\documentstyle[preprint,prb,aps,epsfig,eqsecnum]{revtex}

\newcommand{\be}{\begin{equation}}
\newcommand{\ee}{\end{equation}}
\newcommand{\bn}{\begin{displaymath}}
\newcommand{\en}{\end{displaymath}}
\newcommand{\bs}{\begin{eqnarray}}
\newcommand{\es}{\end{eqnarray}}
\newcommand{\nn}{\nonumber}
\newcommand{\Pf}{\textnormal{Pf}\;}
\newcommand{\up}{\mathsf{up}}
\newcommand{\ri}{\mathsf{right}}
\newcommand{\dw}{\mathsf{down}}
\newcommand{\lf}{\mathsf{left}}
\newcommand{\stk}[1]{\stackrel{#1}{\longrightarrow}}

\begin{document}

\tightenlines
\draft

\title{The Ising Model on Curved Lattices\footnote{Presented at the Workshop on Random Geometry, Krakow 2003}}

\author{Ruben Costa-Santos
\footnote{e-mail R.A.Costa-Santos@phys.uu.nl}}
\address{Spinoza Institute, Utrecht University, Leuvenlaan 4, 3584 CE Utrecht}

\preprint{SPIN 2003-32 ITP-UU 2003-51}

\maketitle

\begin{abstract}

\vspace{-.5cm}

We review recent results concerning finite size corrections to the Ising model free energy on lattices with non-trivial topology and curvature. From conformal field theory considerations  two distinct universal terms are expected, a logarithmic term  determined by the system curvature and a scale invariant term determined by the system  shape and topology. Both terms have been observed  numerically, using the Kasteleyn Pfaffian method,  for lattices with  topologies ranging from the sphere to that of a genus two surface. The constant term is shown to be expressible in terms of Riemann theta functions while the logarithmic correction reproduces the theoretical prediction by Cardy and Peschel for singular metrics.

\end{abstract}

\vspace{8cm}

\section{Introduction}\label{section0}

Finite size corrections provide the shortest bridge between the critical behavior of a lattice model and the conformal field theory description of its universality class. The simplest example of such a connection being  the dependence of finite size corrections on the model central charge for a strip geometry \cite{cardy1}.
In more general geometries, the universal part of the free energy will  depend sensitively on the lattice shape, topology and curvature. For a two dimensional system, characterized by a length scale $L$, the free energy at criticality and for fixed shape has   a large $L$ expansion \cite{cardy2} of the form
\be
      F=  f_0  L^2 + f_b L + C \ln{L} + D +o(1)\label{fexp}.
\ee
where $C$ and $D$ are universal, the logarithmic term coefficient $C$ being determined by the system curvature and the constant term $D$ by its topology and shape.

\begin{figure}[t]
\includegraphics[width=.8\linewidth]{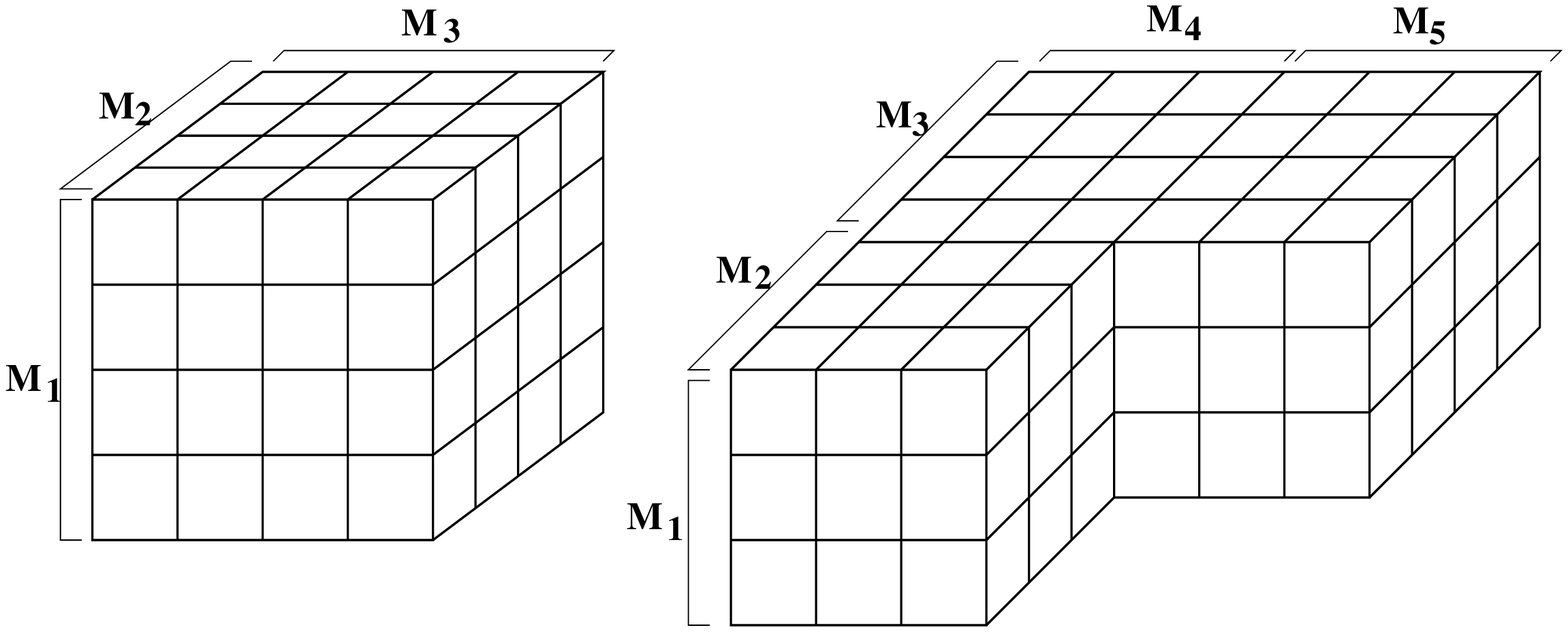}
\vskip 1cm
\caption{Two lattices with spherical topology: the cube and the L-shaped lattice, characterized by integer sizes $M_i$. Conical singularities are related with the corners where three or five faces meet.}
\label{fig1}
\end{figure}

\begin{figure}[b]
\includegraphics[width=.8\linewidth]{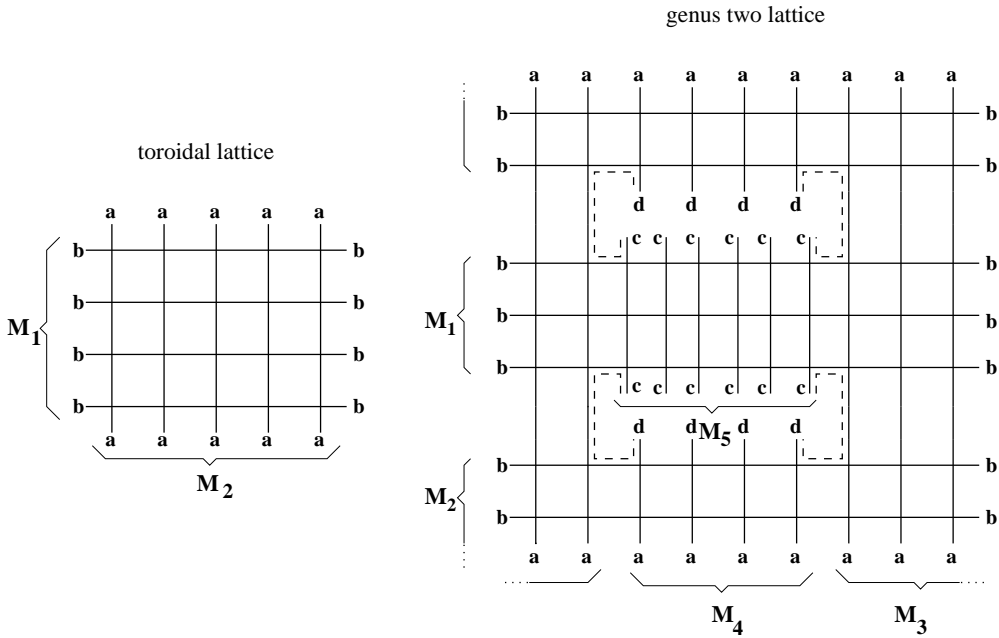}
\vskip 1cm
\caption{The genus two lattice can be seen has a toroidal lattice with an additional handle in the bulk. The boundary edge identifications are given by the letters. The two octagonal faces are marked in dashed line.}
\label{fig2}
\end{figure}

In a series of recent papers \cite{me1,me2,me3}  the author and collaborator have studied these terms for the Ising model on lattices with topologies ranging from the sphere to that of a genus two surface. In this talk we will review these results using as examples the square lattice with spherical topology, Fig. \ref{fig1}, and the square lattice embedded on a genus two surface, Fig. \ref{fig2}.

While it would be desirable to explicitly evaluate $C$ and $D$ in the lattices scaling limit, progress along this direction has been hindered by difficulties in evaluating the free energy. Closed form expressions for the Ising model free energy are known only for a small number of finite size lattices: the torus \cite{kauf}, the cylinder  \cite{mccoy} and some non-orientable \cite{luwu2} geometries, where translational invariance permits the use of Fourier transforms.  
Progress can be made however by using the Kasteleyn  formalism to express the partition function  as a linear combination of determinants of adjacency matrices. These determinants  can then be evaluated numerically for large lattice sizes and arbitrary lattice geometry.

The Ising model on lattices with the topology of the sphere has been  studied previously \cite{gonzalez00,gonzalez0,hlang} and the coefficient  $C$ was obtained for a restricted class of hexagonal lattices \cite{gonzalez}. Topology dependent terms are also observed, for spherical topology, in the context of percolation \cite{ziff}.

The paper is organized in the following way: 

In Section \ref{section1} we review the Kasteleyn method for the evaluation of the Ising model partition function on general two dimensional lattices.   

In Section \ref{section2} we discuss the  constant term  $D$ for lattices embedded on a genus $g$ surface.  Careful numerical studies showed that $D$ can be expressed  in terms of Riemann theta functions, defined below in the text, with half-integer characteristic $k$
\begin{equation}
         D \simeq  \tilde D\  \sum_{k=1}^{4^g}\left| \Theta[k](0|\Omega)\right| . \label{dd}
\end{equation}
This result is reminiscent of multi-loop calculations in string theory \cite{alvarez1,verver1} with $D$  reproducing the modular invariant partition function of the $c=\frac{1}{2}$  conformal field theory on a genus $g$ surface. 
The  lattice shape and coupling constants determine the $g\times g$ matrix $\Omega$. This matrix  is the period matrix of a Riemann surface related with the lattice continuum limit. In Section \ref{section3} we study the  dependence of $\Omega$ on the lattice shape using a discrete formulation of Riemann surface theory.

  In Section \ref{section4} we discuss the logarithmic term $C$ for both spherical  and higher genus lattices. From a conformal field theory point of view it is known \cite{cardy2} that this term is sensitive to the smoothness of the system. For a smooth system it is related with the total curvature by
\be 
      C= -\frac{1}{6} \,c\, \chi \label{33} 
\ee
with $\chi= 2-2g$ being the system Euler characteristic and $c$ the central charge. By opposition, the contribution from a conical singularity with deficit angle $\epsilon=2\pi-\theta$ is
\begin{equation}
     C_\theta= \frac{c\,\theta}{24\pi} \left(1- (2\pi/\theta)^2 \right) \label{22}
\end{equation}
and not a term proportional to $\epsilon$ as one would expect from a delta function in the curvature. For the  scaling limit of a specific lattice the distinction between the two cases depends on the order in which the limit of singular curvature and the thermodynamic limit are taken. We will show that, for the lattices of Figs. \ref{fig1} and Fig. \ref{fig2}, $C$ is given by the sum over all  conical singularities   in the lattice of the  contribution (\ref{22}). Similar results have been reported by Gonzalez \cite{gonzalez}. 

Finally in  Section \ref{section5} we present our conclusions and discuss open problems and avenues for future research.

\section{The Kasteleyn formalism }\label{section1}

Consider a lattice, or more properly a graph,  $G$ drawn without superposition of edges on a surface of genus $g$. Assigning a coupling constant to each edge  and placing an Ising spin at each vertex we can defined the Ising model on $G$.
Kasteleyn  \cite{kast3,russ} showed that its partition function  can be expressed as a linear combination of the Pfaffians of $4^g$ antisymmetric matrices.  This result provides the starting point for the study of the Ising model on non-trivial geometries.  In this section we provide a brief description of the Kasteleyn method, for further details and proofs  we refer the interested reader to references  \cite{me1,kast3,russ}. 

\begin{figure}[h]
\includegraphics[width=.8\linewidth]{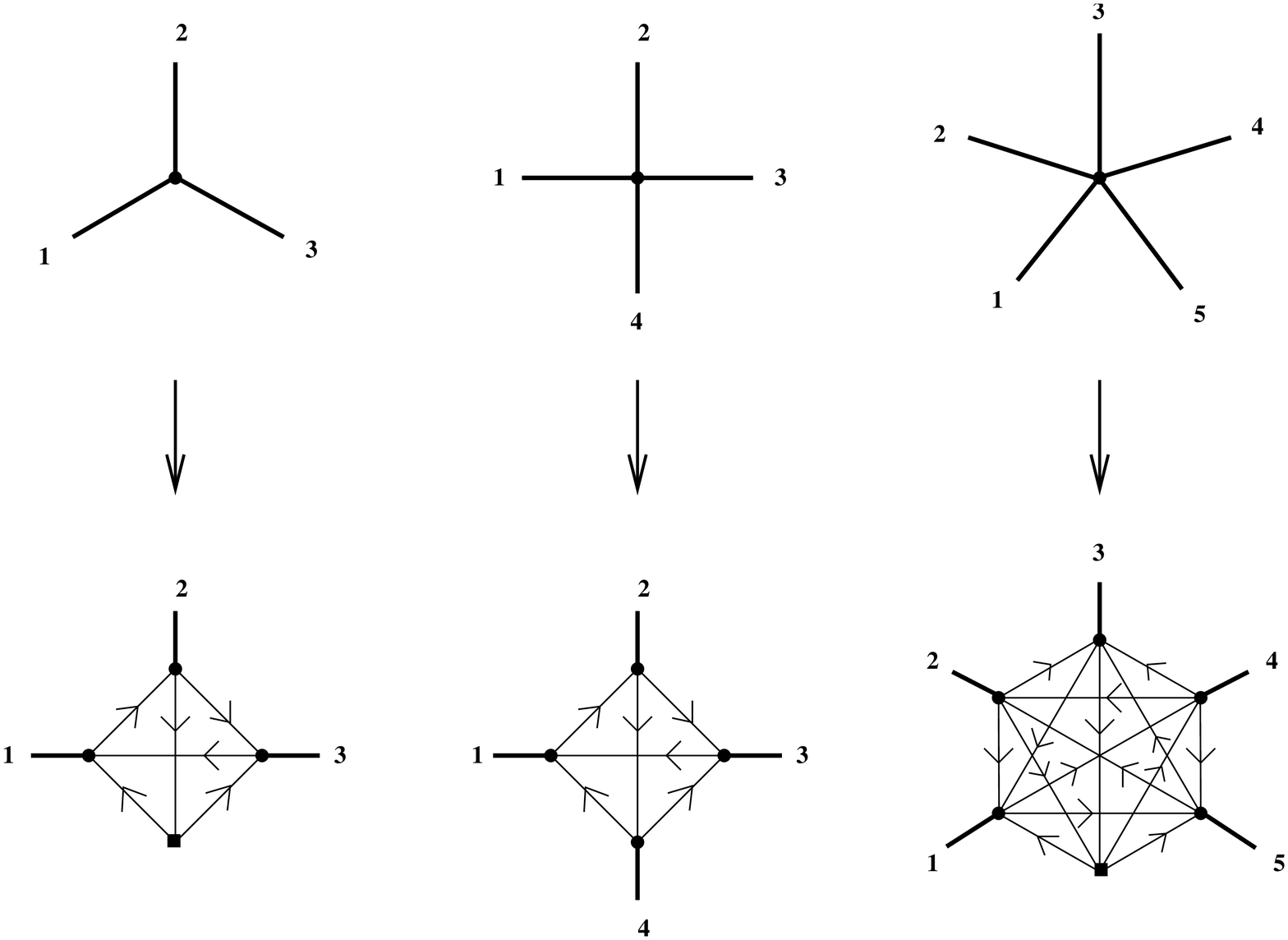}
\vskip 1cm
\caption{The vertex decorations for vertices with three, four and five nearest neighbors.}
\label{fig3}
\end{figure}

Given a  graph embedded on a genus $g$ surface, the Kasteleyn method to obtain the Ising model partition function consists of essentially three steps:

First step: replace each vertex of the original lattice by a cluster of vertices, called the decoration graph, that depends on the vertex coordination number. A possible choice of the decoration graph for vertices with three, four and five nearest neighbors is shown in Fig. \ref{fig3}. The order by which the exterior edges connect to the decoration is not important, meaning that the decoration graphs can be rotated.  
 The resulting decorated lattice will be denoted by $G_d$. Its edges  are assigned  weights: $1$ to the internal edges of the decorations and a weight $w=\tanh{K}$ to the edges inherited from the lattice $G$, with $K$ being the Ising coupling constant on that edge in units of $K_B T$. 

Second step:   orientate the edges of $G_d$ by assigning to each edge a direction, represented  graphically by an arrow. The internal edges of the decorations  have already an orientation as given in Fig. \ref{fig3}. The remaining edges, inherited from the original lattice $G$, are oriented according to the Kasteleyn rule: in such a way that all lattice faces have an odd number of clockwise oriented edges.
An example of an orientation satisfying the Kasteleyn condition is shown in Fig. \ref{fig4} for a genus two lattice.

For any lattice $G_d$ there are many different edge orientations that satisfying  the Kasteleyn condition.  Two lattice edge orientations  are said to be equivalent if they can be related one to the other by a series of arrow reversals,  in which the orientations of all the edges meeting at a given vertex are reversed.  It can be shown \cite{me1,kast3,russ} that for a genus $g$ lattice there are precisely $4^g$ un-equivalent Kasteleyn orientations,  matching the number of spin structures on the corresponding continuum free fermion description. Let  $ \tilde a_i, \tilde b_i$ with $i=1,\ldots,g$ be closed paths on the dual lattice forming a canonical basis of the embedding surface first homology group; then the $4^g$ un-equivalent Kasteleyn orientations can be generated from an initial Kasteleyn orientation by reversing the orientation of the edges crossed by a choice of such loops.

Third step: label the vertices of the decorated graph $G_d$ with an integer from 1 to ${\cal M}$. To each  Kasteleyn edge orientation  associate a ${\cal M}\times {\cal M}$ adjacency matrix  $A$  with entries $A_{ij}$ that vanish if vertices $i$ and $j$ are not connected by an edge and take a value $\pm z$ if  vertices $i$ and $j$ are connected by an edge of weight $z$, the sign being determined by the edge orientation. Schematically 
\be
       A_{ij}= \left\{
              \begin{array}{rl}
              z &  \textnormal{ if    } i \stk{z} j  \\
             -z &  \textnormal{ if    } j \stk{z} i  \\
              0 & \textnormal{ otherwise}
               \end{array}   \right. .
\ee

The Ising model partition function on the graph $G$ is then given by
\be
  Z(K) = 2^{N_V} (\cosh{K})^{N_V} \;\frac{1}{2^g}\; \sum_{i=1}^{4^g} \alpha_i \, \Pf A_i(K)  \label{zz}
\ee
were $N_V$ and $N_E$ are the number of vertices and edges in $G$ and we assume that all edges have the same coupling constant $K$. The sum runs over representatives of the  $4^g$  un-equivalent Kasteleyn orientations. The $\alpha_i$ take values $\pm 1$ and are completely determined by the arrow parity of non-trivial topology loops $ a_i, b_i$  along lattice edges, see \cite{me1} for details. 

This completes our description of the Kasteleyn method. We have reduced the evaluation of the Ising model partition function  to the calculation of determinants of adjacency matrices, using that $\Pf A_i=\sqrt{\det{A_i}}$ for an antisymmetric matrix.

\section{The constant term $D$ on higher genus lattices} \label{section2}

Consider the genus two lattice shown in Fig. \ref{fig2}, it can be seen as a torus with an additional handle on the bulk and it is characterized by five integer sizes $M_i$. Locally it is equivalent to the flat square lattice except around the two octagonal faces. These faces in the continuum limit will correspond to conical singularities on a otherwise flat system. 
To study the finite size corrections we evaluate the free energy on a sequence of lattices with fixed shape and increasing size. This is done by taking lattice  dimensions of the form  $M_i=m_i\, L$ with fixed $m_i$ and increasing $L$. The coupling constants are fixed for all edges to the square lattice isotropic critical value 
\be
\sinh{2K_c}=1. \label{crit}
\ee

Following the discussion of the previous section, the partition function on a genus two lattice can be evaluated in terms of sixteen determinants of adjacency matrices. These matrices and the associated Kasteleyn edge orientations can be labeled as $A(n_{\tilde a_1},n_{\tilde b_1},n_{\tilde a_2},n_{\tilde b_2})$ with $n_x=0,1$.  The starting edge orientation $A(0000)$ is shown in Fig. \ref{fig4} and  an orientation with $n_x=1$ is obtained from the corresponding orientation with $n_x=0$ by reversing the orientation of all the edges crossed by the cycle $x$. 
To  make connection with the theta function  characteristics and allow for more compact equations we will also use the alternative notations
\be
   A(n_{\tilde a_1},n_{\tilde b_1},n_{\tilde a_2},n_{\tilde b_2})= A\left[\begin{array}{cc}   n_{\tilde b_1} & n_{\tilde b_2} \\ n_{\tilde a_1} & n_{\tilde a_2} \end{array} \right]=A_i \label{orient}
\ee 
 with the integer label given by $i={\textstyle 16-8n_{\tilde a_1}- 4n_{\tilde b_1}-2n_{\tilde a_2}-n_{\tilde b_2}}$.

In terms of these orientations, the Ising model partition function on the genus two lattice is  given by
\bs
Z=\frac{\alpha_0}{4}(P_1-P_2-P_3-P_4-P_5+P_6+P_7+P_8- P_{9}  \nn \\  
    +P_{10}+P_{11}+P_{12}-P_{13}+P_{14}+P_{15}+P_{16})  \label{zzz}
\es 
where $P_i=\Pf A_i$ and $\alpha_0$ is the size dependent pre-factor in (\ref{zz}). Closed form expressions for these Pfaffians are  difficult to obtain, since  the lattice  is not translationally invariant, but they can be evaluated numerically for relatively large lattice sizes. 

To factor out the bulk term $f_0$ and the logarithmic correction $C$  we can, as in the toroidal case \cite{ferdfish}, consider ratios of  determinants of adjacency matrices  to the largest among the sixteen.
These ratios are found to converge to well defined values for large lattice size $N_V$, see  Fig. \ref{fig5} for an example. More precisely, we find that in the  $L\rightarrow \infty$  limit the ratios of determinants satisfy at criticality
\be
  R_i\equiv \left. \frac{\det\left( A{\scriptsize \left[\begin{array}{cc} c_1&c_2\\c_3&c_4\end{array} \right]}\right)}
             {\det \left(A\scriptsize{\left[\begin{array}{cc}0 & 0\\ 0 & 0 \end{array} \right]}\right)} \,\right|_{T_c} =
  \left|\frac{\Theta\scriptsize{\left[\begin{array}{cc} c_1/2&c_2/2\\c_3/2&c_4/2  \end{array} \right]}\left(0,\Omega \right)}
             {\Theta\scriptsize{\left[\begin{array}{cc}  0&0\\ 0&0 \end{array} \right]}\left(0,\Omega \right)}\right|^2 \label{resultati}
\ee
for the 16 combinations of $c_x=0,1$ and with the label $i$  defined below equation (\ref{orient}). 
The  genus two Riemann theta functions are defined by  the quickly converging series 
\be
   \Theta {\left[\begin{array}{c}  { \mbox{\boldmath $\alpha$}} \\  { \mbox{\boldmath $\beta$}} \end{array} \right]}  \left({\bf z},\Omega \right)
   =\sum_{{\bf n}\in Z^2} \exp{\left[i\pi({\bf n}+ { \mbox{\boldmath $\alpha$}})^T \Omega({\bf n}+{ \mbox{\boldmath $\alpha$}})+2\pi i({\bf n}+{ \mbox{\boldmath $\alpha$}})^T ({\bf z}+{ \mbox{\boldmath $\beta$}})\right]}   \label{theta}
\ee
where {\boldmath $\alpha$}, {\boldmath  $\beta$}, ${\bf  z}$ and ${\bf n}$ are  2-vectors half-integers, complex numbers and integers respectively. Six of the sixteen theta functions in (\ref{resultati})  are odd functions of ${\bf z}$ and therefore vanish at the origin. The corresponding ratios of determinants are found to vanish in the large $N_V$ limit.

The $2\times 2$ period matrix $\Omega$  is a symmetric, complex valued matrix with a positive definite imaginary part. It is determined by the lattice shape and coupling constants, its value in (\ref{resultati}) can be obtained directly by numerically fitting the ratios of determinants or from first principles, as we will see in the next section, by using discrete holomorphy methods. For locally square lattices, $\Omega$  is found to be purely imaginary 
\be
     \Omega=i\left[\begin{array}{cc} \Omega_{11} &\Omega_{12}\\\Omega_{12}& \Omega_{22}\end{array} \right],
\ee
this is not the case for triangular lattices where  the period matrix is in general complex \cite{me2}. The same property is observed on toroidal lattices where the modular parameter  is imaginary for squared lattices \cite{ferdfish} and complex for triangular lattices \cite{nashoco}.

In Table \ref{table1} these results are illustrated for the lattice with shape $(m_i)=(1,1,1,1,1)$. The fifteen determinant ratios are shown, both the largest lattice size evaluation and the $L\rightarrow \infty$ extrapolation obtained using a fit with a quadratic polynomial in $1/N_V$. These ratios are compared with ratios of theta functions 
\be
     \Theta_{(16-8d_1-4c_1-2d_2-c_2)}(\Omega)=\Theta\scriptsize{ \left[\begin{array}{cc} c_1/2& c_2/2\\ d_1/2& d_2/2\end{array} \right]}\left(0,\Omega \right)/\Theta{\scriptsize \left[\begin{array}{cc} 0&0\\ 0&0\end{array} \right]}\left(0,\Omega \right) \label{thethe}
\ee
with a period matrix obtained by a suitable numerical fitting procedure. The precision to which the two sets of numbers agree is remarkable, typically a precision from $10^{-4}$ to $10^{-6}$.

\section{Discrete holomorphy and the period matrix}\label{section3}

The  matrix $\Omega$ in Eq. (\ref{resultati}) can be seen as the period matrix of a Riemann surface related with the lattice continuum limit. In this section we study the dependence of this matrix  on the lattice shape using a discrete formulation of holomorphy. 
The basic idea is to formulate Riemann surface theory on a discrete setting by using finite difference operators \cite{me1,me2,mercat,mercat3}. These operators act on  quantities defined on the lattice p-elements: vertices, oriented edges and faces that we will call respectively the lattice functions, differentials and volume forms.

A lattice function $f$ is determined by its value on the lattice vertices and can be represented by a  $N_V$-vector   $f[n]: n= 1,\ldots, N_V$ after an integer labeling of the lattice $N_V$ vertices is chosen.

 A lattice differential $w$ is determined by its value on the lattice oriented edges. Referring to a fixed drawing of the lattice, we define horizontal edges to be oriented from left to right and vertical edges from bottom to top. A lattice differential is then represented  by a  $N_V\times 2$-matrix  $w[n|p]: n=1,\ldots,N_V; \ p=1,2$ where $[n|1]$ stands for the horizontal edge immediately right of vertex $n$ and $[n|2]$ for the vertical edge immediately above $n$, see Fig. \ref{fig7}. The integral of a lattice differential $w$ along a path of lattice edges $C$ is the sum of the values that $w$ takes on the edges included in $C$
 \be
   \int_{C} w= \sum_{[n|p]\in C } \pm \,w[n|p]
\ee
with a minus sign for edges with opposite orientation to that of the path.

A lattice volume form $\eta$ takes values on the lattice faces and is represented by a $(N_V-2)$-vector  $\eta [q]: q= 1,\ldots,N_V-2$ after a labeling of the  $N_V-2$ lattice faces is chosen. The integral of a lattice volume form $\eta$  over a given lattice area $A$ is the sum of the values that $\eta$ takes on all faces included in $A$.

It is important to relate the labeling of vertices, edges and faces. Vertices and edges are already related by the labeling introduced above, to relate vertices with faces we introduce some additional notation: if $q$ is a lattice squared face then  $\hat q_1$ is its lower left vertex,  if $q$ is an octagonal face then $\hat q_1$ and $\hat q_2$ correspond to the two vertices that can be seen as its  lower left vertices, see Fig. \ref{fig2}. Conversely for a given vertex $n$ we denote by $\tilde n$ the lattice face of which $n$ is a lower left vertex.  

We now define finite difference operators acting on these quantities, which are the lattice versions of the exterior derivative, the co-derivative  and the Hodge star operator, see Fig. \ref{fig55}.

The  lattice exterior derivative  $d$ is a linear operator defined by
\bs
  (d\, f)\, [n|1]&=& f[\ri(n)] - f[n] \label{der}\\
  (d\, f)\, [n|2]&=& f[\up(n)] - f[n] \nn\\
 (d\, w)\, [q]&=& \sum_i \left( w[\hat q_i|1] + w[\ri(\hat{q}_i)|2] - w[\up(\hat{q}_i)|1] - w[\hat{q}_i|2]\right) \nn
\es
with $i$=1,2 for octagonal faces and $i$=1 for squared faces. The functions $\ri(n), \lf(n), \up(n)$ and $\dw(n)$  give the label of the vertex immediately right, left, above and below of the vertex $n$.

The  lattice co-derivative $\delta$ is the operator defined as
\bs
   (\delta\,w)\,[n]&=&  w[\lf(n)|1]-w[n|1] +  w[\dw(n)|2]-w[n|2]\label{del} \\
  (\delta\,\eta\,)\,[n|1]&=&  \eta[\tilde n]-\eta[\widetilde{\dw}(n)]\nn \\
  (\delta\,\eta)\,[n|2]&=&  \eta[\widetilde{\lf}(n)]-\eta[\tilde n] \nn \\
\es
where $\widetilde{\lf}(n)$ stands for the face given by the tilde of the vertex ${\lf}(n)$ and similarly for $\widetilde{\dw}(n)$.

The  discrete Hodge star acting on lattice differentials is defined as
\bs
  (\star w)[n|1] &=& -w[\dw(n)|2]  \label{star}\\
  (\star w)[n|2] &=& \ \ w[\lf(n)|1] \nn .
\es

These discrete operators are defined in such a way that they satisfy most of the usual properties of their continuum counterparts: the lattice exterior derivative satisfies a discrete version of Stokes theorem; we have that  $dd=\delta\delta=0$ and the two operators are adjoint of each other under the trivial inner product of lattice functions, differentials and volume forms. These properties are exact for all lattices at finite size.  Such is not the case for some important properties of the discrete Hodge star operator that, as we will see below,  are only satisfied in the limit of large lattice size.

In exact analogy with the continuum definitions, a  lattice differential is said to be  harmonic if it is both closed and co-closed,
\be
     w \textnormal{   is harmonic } \equiv \left\{
              \begin{array}{cl}
                    (d\, w)[q] =0 ,   & \ q=1,\ldots,N_F\\ 
                (\delta\, w)[n]=0 ,   & \ n=1,\ldots,N_V 
              \end{array}   \right. , \label{harm}
\ee
where the conditions must be met for all $N_F$ lattice faces  and all  $N_V$ lattice vertices.
In general, on a genus $g$ lattice, the system of equations (\ref{harm}) has  $2g$ linearly independent solutions. Notice that the number of of unknowns $w[n|p]$ minus the number of equations is already
\be
    N_E-N_V-N_F= -\chi =2g-2 
\ee
where $N_F$ is the number of lattice faces and $\chi$ is the Euler characteristic of the lattice. Two additional solutions are provided by constant differentials with independent vertical and horizontal components. That these are the only solutions follows from the fact that a harmonic differential, discrete or continuum, is completely determined by its periods along the lattice, or surface, non-trivial loops and there are only $2g$ such loops in a genus $g$ lattice.

Finally a lattice differential is said to be holomorphic if it is harmonic and it satisfies $\star w= -i w$. We  then need to require the lattice Hodge star to be an endomorphism on the space of  harmonic differentials and to  satisfy $\star\star=-1$. Unfortunately these two properties are only approximately verified on a finite size lattice, see \cite{me1} for details. Except for some specially symmetric lattices, one expects these properties to be exact only in the large $N_V$ limit.

As in continuum Riemann Surface theory, the lattice period matrix can be defined it terms of the lattice holomorphic differentials.  We start by  solving the linear system of equations (\ref{harm}) for the lattice  harmonic differentials. There will be $2g$ independent solutions spanning a $2g$ linear space that  can be decomposed, almost precisely, in two sub-spaces of holomorphic and anti-holomorphic lattice differentials by means of the projection operators 
\be 
P=(1 + i\star)/2 \ \ \textnormal{and} \ \ \bar P=(1 - i\star)/2 \label{proj}.
\ee
We then look for a basis of the holomorphic sub-space   \mbox{$\{\Gamma_k : k=1\ldots,g\}$} with normalization
\bs
    \int_{a_k} \Gamma_l &=& \delta_{kl} \label{holo} \\
     \int_{b_k}   \Gamma_l &=& \Omega_{kl} \label{period}
\es
where the integrals are along a choice  $a_j, b_j$ of closed paths of lattice edges representing a basis of the embedding surface first homology group.

Equation (\ref{period}) gives our evaluation of the finite size period matrix.
 Detailed numerical studies \cite{me1,me2} showed the resulting matrix  to be a complex, symmetric  matrix  with a positive definite imaginary part. For any  lattice of the form shown in Fig. \ref{fig2}, the fixed shape $N_V\rightarrow \infty$ limit of this matrix reproduces with excellent precision the period matrix $\Omega$ in equation (\ref{resultati}). The lattice period matrix depends however on the choice of the first homology group basis  but different period matrices for the same lattice will be related by a modular transformation on the  limit of large lattice size.

\begin{figure}[t]
\includegraphics[width=.8\linewidth]{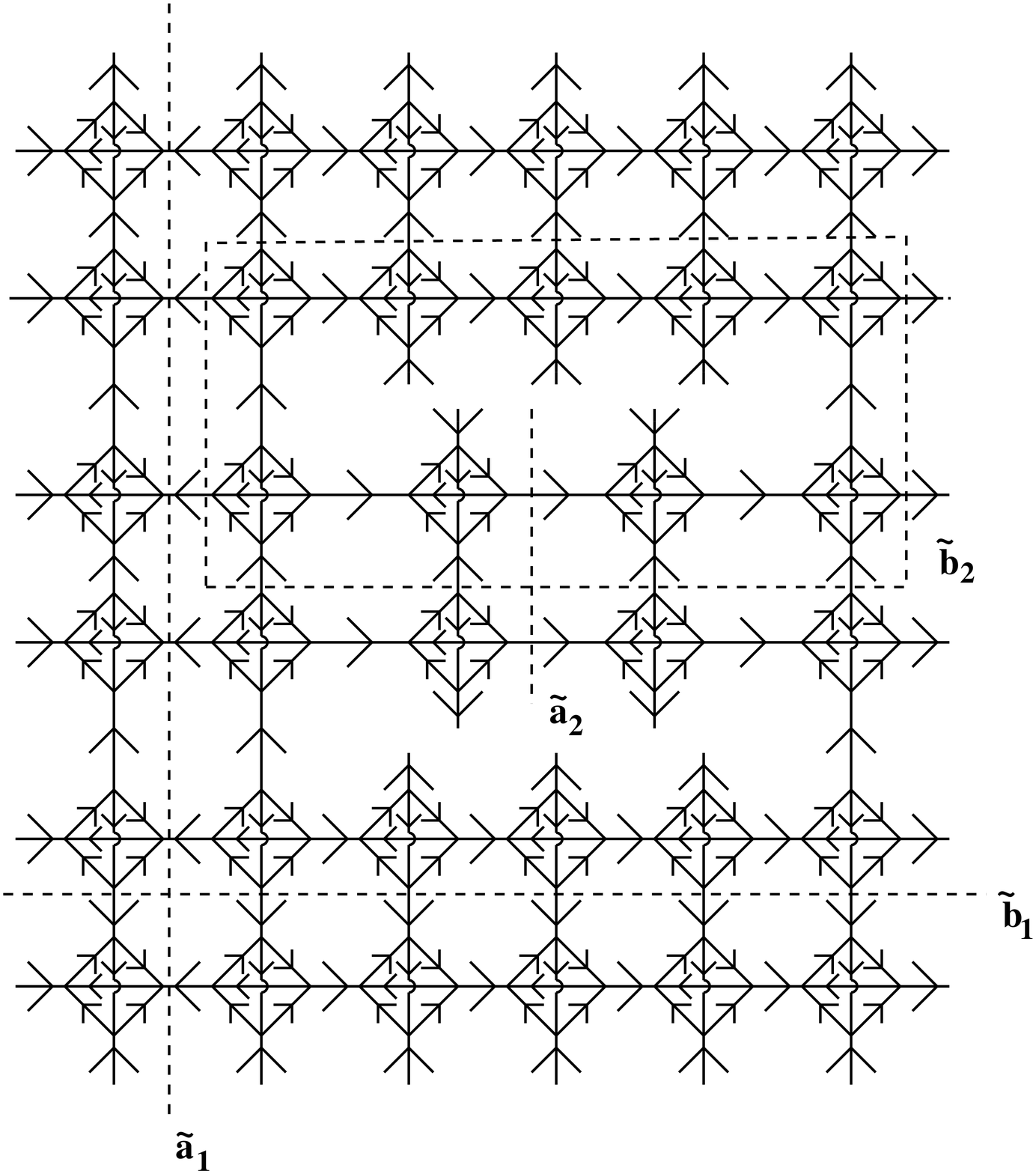}
\vskip 1cm
\caption{The $A(0000)$ Kasteleyn edge orientation of the decorated genus two lattice with integer sizes $(M_i)=(2,4,3,3,2)$.The remaining fifteen Kasteleyn orientations are obtained from this one by reversing the orientations of all edges crossed by a selection of the $\tilde a_i,\tilde b_i$ loops. }
\label{fig4}
\end{figure}

\begin{figure}[h]
\includegraphics[width=.8\linewidth]{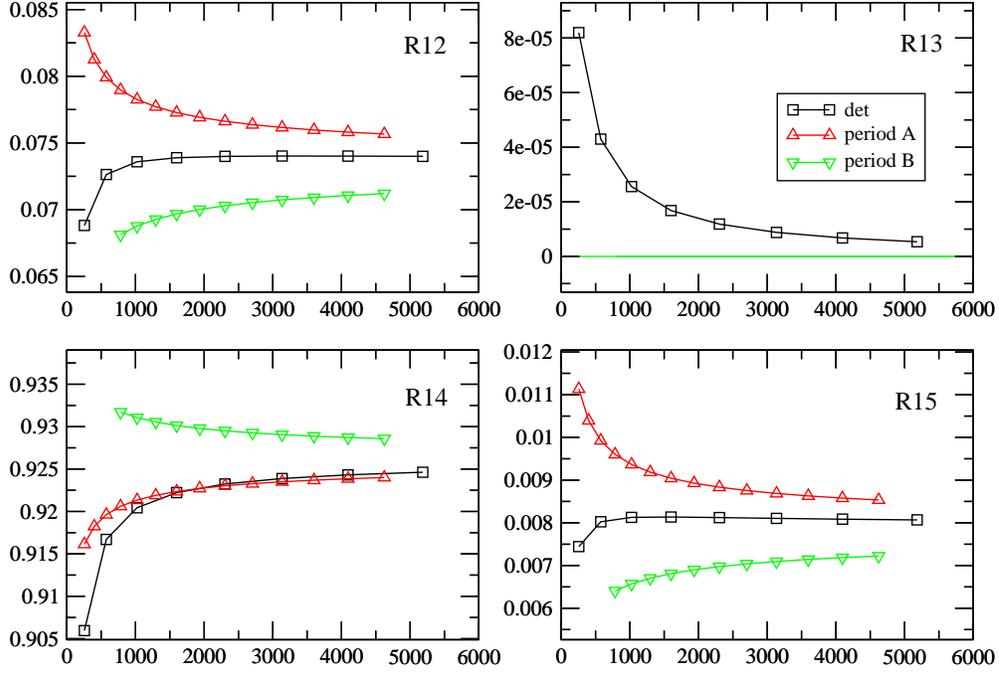}
\vskip 1cm
\caption{Ratios of determinants $R_i^2$ for the genus two lattice with shape $(m_i)=(11111)$ in function of the number of lattice points $N$  and the theta function ratios $\theta_i^2$ for period matrices evaluated using  two different first homology group basis, A and B. The ratio $R13$ corresponds to an odd characteristic theta function and vanishes in the large $N_V$ limit.}
\label{fig5}
\end{figure}

\begin{figure}[h]
\includegraphics[width=.8\linewidth]{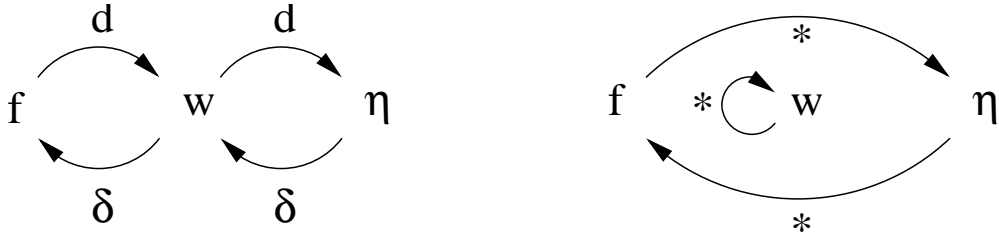}
\vskip 1cm
\caption{The lattice functions $f$, differentials $w$ and volume forms $\eta$ and the finite difference operators acting on them.}
\label{fig55}
\end{figure}

\begin{figure}[t]
\includegraphics[width=.8\linewidth]{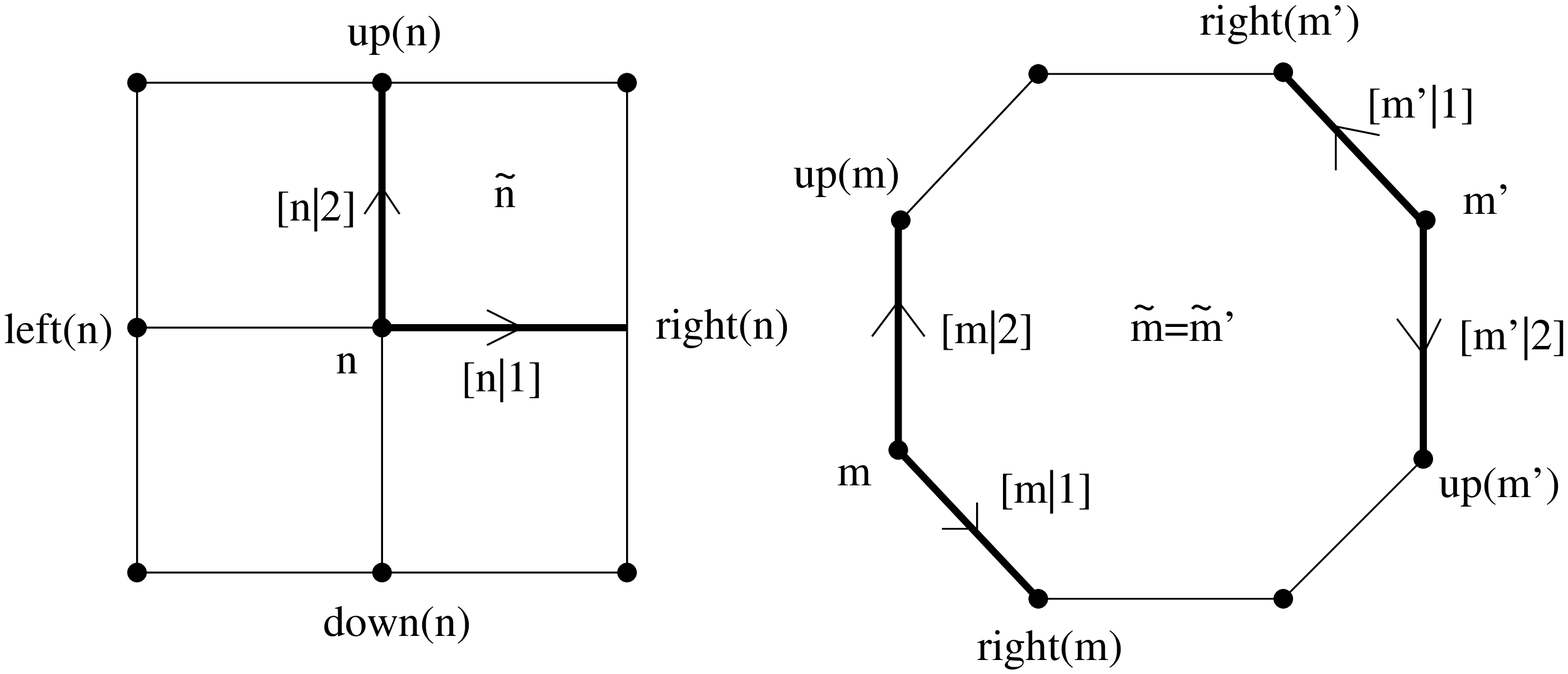}
\vskip 1cm
\caption{The labeling of vertices, edges and faces for squared and octagonal faces.}
\label{fig7}
\end{figure}

In  Fig. \ref{fig5} four ratios of determinants of adjacency matrices are plotted, in function 
of the lattice size $N_V$, together with theta function ratios (\ref{thethe}) for period matrices evaluated with two different basis of the first homology group. The basis A is similar to the one shown in Fig. \ref{fig4} while the basis B has the $a_i$ and $b_i$ loops interchanged from  Fig. \ref{fig4}. We see that, for each ratio, the three values converge to a common value on the large $N_V$  limit.
In Table \ref{table1}  numerical values are given for the  determinant
  and  theta function ratios. The theta function ratios are shown both for a  period matrix fitted numerically to the ratios of determinants and the $L\rightarrow \infty$ extrapolated lattice  period matrix evaluated using the procedure described in this section. Numerical values for the entries of the two period matrices agree within $1\%$, the difference being  due to the difficulty in the obtaining the  $L\rightarrow \infty$ extrapolations.

The method introduced in this section to evaluate lattice period matrices can be generalized to the anisotropic squared lattice \cite{me1} and the general triangular lattice \cite{me2}.

\section{The logarithmic term $C$}\label{section4}

We now consider the logarithmic  term   for both   the spherical lattices of Fig. \ref{fig1} and the genus two lattice of Fig.  \ref{fig2}. The toroidal lattice will also be considered for illustration purposes. 
 The Ising model free energy can be  expressed  in terms of the determinants of one, four and sixteen  adjacency matrices for the spherical, toroidal and genus two lattice respectively. Again we will consider sequences of lattices with fixed shape $m_i$ and increasing size $M_i=m_i L$ at the isotropic critical coupling (\ref{crit}).
For the higher genus lattices we learned from (\ref{resultati}) that the different adjacency matrix determinants converge rapidly to a common bulk term times a topology and shape determined factor,
\be
       \Pf A_i (N_V)= \Theta_i \ \tilde{Z}(N_V)
\ee
where the $ \Theta_i$ are constants for large enough lattice size $N_V$. It follows that since the logarithmic correction is entirely due to the $\tilde{Z}$ term,
\be
   - F=  \ln{\left(\frac{\alpha_0}{2^g} \sum_{i=1}^{4^g} \alpha_i \Theta_i \right)} + \ln{\tilde{Z}},
\ee
it can be equivalently evaluated from any non-vanishing Pfaffian. For this purpose it is convenient to introduce the auxiliary quantities
\be
   F_i(N_V)\equiv - \ln\left({\frac{\alpha_0}{2^g}\alpha_i\Pf A_i}\right)= F(N_V)  +  \ln{\left( \sum_{k=1}^{4^g} \frac{\alpha_k \Theta_k}{\alpha_i \Theta_i} \right)} 
\ee
where $i$ labels one of the $4^g$ un-equivalent Kasteleyn orientations. In the  thermodynamic limit $F_i$ differs from $F$ only by a constant.

\begin{figure}[t]
\includegraphics[width=.8\linewidth]{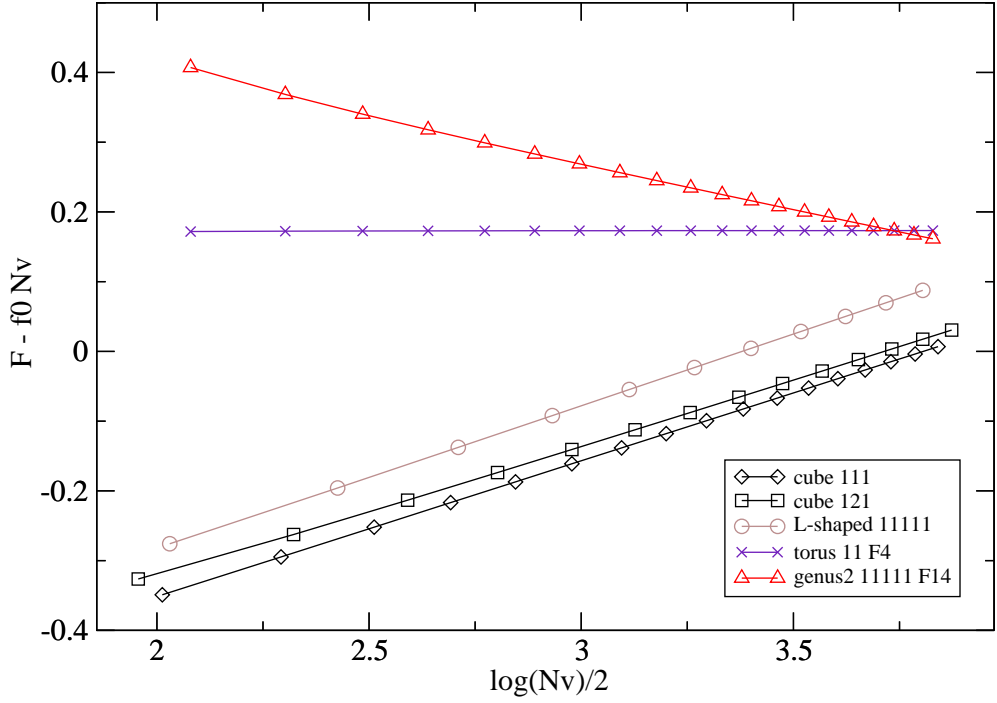}
\vskip 1cm
\caption{The residual free energy in function of the logarithm of the lattice size $N_V$ for various lattices. For the torus and genus two lattice only one Pfaffian term  $F_i$ is considered. The uncertainties on the values are smaller than the symbols used.}
\label{fig6}
\end{figure}

The logarithmic correction $C$ can be found  by subtracting  to the lattice free energy the leading volume term $f_0$, that is known \cite{ferdfish} to be 
\be
    f_0= 2G/\pi + \frac{1}{2} \ln{2}
\ee  
for the squared lattice, with $G$ being the Catalan constant. 
In Fig. \ref{fig6}  the residual free energy $F - f_0 N_V$ is plotted as function of the logarithm of the lattice size, $(\log{N_V})/2$, for lattices with the various geometries. A clear linear behavior is observed and the difference between positive, zero and negative curvature is patent for the spherical, toroidal and genus two lattice.

For each lattice we evaluate a series of values converging to $C$  by doing linear regressions on sets of four  lattices with consecutive increasing sizes. Examples of values  obtained  are shown in Table \ref{table2}. We find convergence to a well defined value of $C$, the rate of convergence depending on the shape of the lattice, being fastest for the more symmetric lattices. 
The fact that the two spherical lattices, the cube and the L-shaped lattice, have different logarithmic corrections $C$ already points out that this correction cannot be explained by the smooth metric contribution equation (\ref{33}) but it is given  by a sum over the conical singularities contributions of equation (\ref{22}). 
The  $C_\theta$ corrections  for vertices where three, five and eight  squared faces meet are listed on Table \ref{table3}. While the singularities of the genus two lattice are located at the octagonal faces we can by duality see them as vertices where eight faces meet. The total logarithmic correction $C$ is obtained by summing  the $C_\theta$ contributions for all singularities occurring in a lattice.
 As shown in  Table  \ref{table2},  this reproduces the numerical values of $C$ with high accuracy. The validity of equation (\ref{22})  for all the lattices considered is clearly established from the numerical results. 

It should be remarked that the conical singularity behavior is related with  the large scale structure of the lattice and not the  small scale structure.  In \cite{me2} it was shown that a vertex where more, or less, than four faces meet does not  necessarily constitute a conical singularity. Conical singularity behavior on the finite size corrections is observed when a regular lattice is folded at large scale into a conical shape spanning an angle $\theta$.

\section{Conclusions} \label{section5}

In this talk we reviewed recent research on finite size corrections, the universal terms, of the Ising model free energy in lattices with curvature. While these terms are well understood from a conformal field theory point of view they are not usually studied in the scaling limit of specific lattice geometries.
 Using the Kasteleyn method and numerical evaluation of the adjacency matrix determinants it was possible to study a number of different geometries consisting of regular lattices folded around conical singularities.

The scale invariant term $D$ was shown to be a modular invariant quantity expressed in term of Riemann theta functions, the dependence on the lattice shape and coupling constants being realized through a period matrix $\Omega$. This result provides a  lattice regularized picture of the classical winding part of the  partition function in multi-loop calculations in string theory \cite{alvarez1,verver1}.  The study of the exact dependence of the period matrix  on the lattice shape  provides also a testing ground for ideas of discrete holomorphy.

On the other hand the study of the logarithmic term  $C$ gave a first answer to the long standing question, posed by Cardy and Peschel in \cite{cardy2}, to which of the two forms (\ref{33}) or (\ref{22}) for the logarithmic correction is observed in specific lattice models. 
We found that for the regular lattices with conical singularities  considered in this paper, the logarithmic correction is given by a sum over conical singularities contribution of the form (\ref{22}). 

It is natural to ask what kind of lattice would have a logarithmic term given by equation (\ref{33}), the smooth metric contribution. Such a lattice should have its curvature spread over the all lattice with the  local curvature being non-zero everywhere but vanishing in the continuum limit. It is not clear how this can be accomplished or what the criticality condition should be for such a lattice. 
In this paper we have discussed a number of  non-usual lattices boundary conditions but it seems that to obtain (\ref{33}) one must also consider lattices with non-usual structure or a non-usual limiting procedure. One thing is clear, since a continuum result must have a well defined regularization procedure, such lattices must exist.

\bigskip
\centerline{{\bf Acknowledgments}}
\bigskip

 This work was partially supported  by the EU grant HPRN-CT-1999-00161. Most of the work described in this talk was done in collaboration with Prof. Barry McCoy.

\pagebreak

\pagebreak

\pagebreak

\pagebreak

\pagebreak

\pagebreak

\pagebreak

\begin{table} \centering
\begin{tabular}{|c|cc|cc|}
i &   $R^2_i(4N_V=5184)$ &   $R^2_i( N_V\rightarrow \infty)$ &$\theta^4_i$(fit)&$\theta^4_i$(eval)\\ \hline
1  & 0.066121 & 0.066386 & 0.066378 & 0.066794 \\ 
2  & 0.000012 & $1.4\times 10^{-6}$& 0.000000 & 0.000000 \\ 
3  & 0.000009 & $1.3\times 10^{-6}$& 0.000000 & 0.000000 \\
4  & 0.000009 & $1.3\times 10^{-6}$& 0.000000 & 0.000000 \\
5  & 0.000005 & $1.1\times 10^{-7}$& 0.000000 & 0.000000 \\
6  & 0.924624 & 0.925554 & 0.925550 & 0.924957 \\
7  & 0.000288 & 0.000287 & 0.000286 & 0.000301 \\
8  & 0.857835 & 0.859467 & 0.859457 & 0.858464 \\ 
9  & 0.000012 & $1.4\times 10^{-6}$& 0.000000 & 0.000000 \\
10 & 0.066121 & 0.066386 & 0.066378 & 0.066794 \\
11 & 0.074001 & 0.074171 & 0.074164 & 0.074742 \\ 
12 & 0.074001 & 0.074171 & 0.074164 & 0.074742 \\ 
13 & 0.000005 & $1.1\times 10^{-7}$& 0.000000 & 0.000000 \\
14 & 0.924624 & 0.925554 & 0.925550 & 0.924957 \\
15 & 0.008069 & 0.008077 & 0.008072 & 0.008249 \\
16 & 1.000000 & 1.000000 & 1.000000 & 1.000000 \\\hline\hline 
\multicolumn{2}{|c|}{} &$\Omega_{11}$&$\Omega_{12}$&\multicolumn{1}{|c|}{$\Omega_{22}$}\\\hline 
\multicolumn{2}{|c|}{fit value}& 1.704 &  $-$1.408 & \multicolumn{1}{|c|}{ 2.816}  \\ 
\multicolumn{2}{|c|}{evaluated value}& 1.701  & $-$1.403  &\multicolumn{1}{|c|}{ 2.806}\\ 
\end{tabular}
\vskip 1cm
\caption{Comparison between ratios of determinants $R_i$ and ratios of theta functions $\theta_i$  for the genus two lattice with shape $(m_i)=(11111)$. Ratios of determinants are given both for the largest $N_V$ evaluation and for the  $N_V\rightarrow \infty$ extrapolation. The  theta function ratios are for a numerically fitted period matrix and for a period matrix evaluated according to the procedure of Section \ref{section3}.  }
\label{table1}
\end{table}

\begin{table}[h] \centering
\begin{tabular}{|c|cc|c|c|c|}
lattice:& \multicolumn{2}{c|}{cube}& {L-shaped}& torus (++) & genus 2 ($-$$-$+$-$)  \\ \hline
shape:& (111)  & (121)&  (11111) & (11)   &   (11111)  \\ \hline
fits&$-$0.19468& $-$0.19043& $-$0.20531 &  0.00080 &  0.12858  \\
&$-$0.19463& $-$0.19142& $-$0.20573 &  0.00062 &  0.12786  \\
&$-$0.19459& $-$0.19201& $-$0.20571 &  0.00050 &  0.12735 \\
&$-$0.19456& $-$0.19240& $-$0.20567 &  0.00041 &  0.12698 \\\hline 
expected: & \multicolumn{2}{c|}{$-$0.19444}& $-$0.20556& 0& 0.125 \\
\end{tabular}
\vskip 1cm
\caption{The logarithmic correction $C$ for lattices with various topologies. A sequence of fits of four consecutive lattices of increasing size is seen to converge smoothly to the predicted value. For a more complete set of  data see \cite{me3}.}
\label{table2}
\end{table}

\begin{table}[b] \centering
\begin{tabular}{|c|ccc|}
\# squares  &   $\epsilon$    &   $\theta$    &  $C_\theta$\\\hline
   3  &     $\pi/2$  &  $3 \pi/2$   & $-$0.024306\\
   5  &    $-\pi/2$&    $5 \pi/2$  &   0.01875\\
   8  &    $-2\pi$ &    $4 \pi $   &  0.0625\\ 
\end{tabular}
\vskip 1cm
\caption{The deficit angle, spanning angle and conical singularity contribution for vertices with various coordination numbers.}
\label{table3}
\end{table}

\end{document}